\documentclass[12pt,preprint]{aastex}

\newcommand{\be}{\begin{equation}}
\newcommand{\ee}{\end{equation}}
\newcommand{\bea}{\begin{eqnarray}}
\newcommand{\eea}{\end{eqnarray}}

\shorttitle{Benefits of Photometric Follow-Up}
\shortauthors{Col\'on \& Ford}
\begin{document}
\title{Benefits of Ground-Based Photometric Follow-Up for Transiting Extrasolar Planets Discovered with $Kepler$ and CoRoT}

\author{Knicole D.\ Col\'on\altaffilmark{1}, Eric B.\ Ford\altaffilmark{1}}

\altaffiltext{1}{Department of Astronomy, University of Florida, 211 Bryant Space Science Center, P.O. Box 112055, Gainesville, FL 32611-2055, USA}

\begin{abstract}
Currently, over forty transiting planets have been discovered by ground-based photometric surveys, and space-based missions like $Kepler$ and CoRoT are expected to detect hundreds more.  Follow-up photometric observations from the ground will play an important role in constraining both orbital and physical parameters for newly discovered planets, especially those with small radii ($R_p \lesssim$ 4 $R_{\oplus}$) and/or intermediate to long orbital periods ($P \gtrsim$ 30 days).  Here, we simulate transit light curves from $Kepler$-like photometry and ground-based observations in the near-infrared (NIR) to determine how jointly modeling space-based and ground-based light curves can improve measurements of the transit duration and planet-star radius ratio.  We find that adding observations of at least one ground-based transit to space-based observations can significantly improve the accuracy for measuring the transit duration and planet-star radius ratio of small planets ($R_p \lesssim$ 4 $R_{\oplus}$) in long-period ($\sim$1 year) orbits, largely thanks to the reduced effect of limb darkening in the NIR.  We also demonstrate that multiple ground-based observations are needed to gain a substantial improvement in the measurement accuracy for small planets with short orbital periods ($\sim$3 days).  Finally, we consider the role that higher ground-based precisions will play in constraining parameter measurements for typical $Kepler$ targets.  Our results can help inform the priorities of transit follow-up programs (including both primary and secondary transit of planets discovered with $Kepler$ and CoRoT), leading to improved constraints for transit durations, planet sizes, and orbital eccentricities.
\end{abstract}

\section{Introduction}
\label{SecIntro}
Over the past decade, ground-based photometry has discovered over forty planets that transit their host star.  A majority of these known transiting planets have radii and masses that are comparable to Jupiter, and most have orbital periods that are less than five days.\footnote[1]{http://exoplanet.eu/}  Currently, space-based planet searches (e.g., CoRoT and $Kepler$) are poised to extend the reach of the transit method to small planets ($R_p \lesssim$ 4 $R_{\oplus}$) and longer orbital periods ($P \gtrsim$ 30 days).  These systems present exciting opportunities for follow-up observations to characterize the physical properties of potentially rocky planets.  As the number of known transiting planets grows, it will become increasingly important to make the best possible use of follow-up resources (Clarkson et al.\ 2007; O'Donovan et al.\ 2007; Pont et al.\ 2007; Latham et al. 2008).  Indeed, merely confirming the planet candidates from CoRoT and $Kepler$ will present a significant challenge, as the target stars will be fainter, the planets will be smaller, and the orbital periods will be longer than for the targets discovered in ground-based surveys.  High-resolution imaging will be important for rejecting blends with background objects or wide binary companions. Moderate-resolution optical and infrared spectroscopy will be essential for rejecting spectroscopic binaries and hierarchical systems.  Finally, at least several high-precision Doppler observations will be needed to confirm planet candidates and to measure planet masses and orbits (Gautier et al.\ 2007; Latham 2007; Brown \& Latham 2008).

There will continue to be an important role for follow-up photometric observations.  For ground-based transit searches, follow-up observations have routinely obtained higher quality photometry, provided more precise measurements of planet and orbital parameters, and served as the basis for transit timing variations.  In this paper, we focus on yet another potential role for follow-up photometric observations.  Due to an approximate degeneracy between the stellar limb darkening parameters and the impact parameter (the minimum distance from the center of the planet to the center of the stars when projected onto the sky plane; e.g., see also P\'al 2008), white light observations provide only weak constraints on the impact parameter.  The uncertainty in the impact parameter propagates to interfere with the measurement of the planet-star radius ratio and transit duration (Knutson et al.\ 2007).  An accurate measurement of the radius ratio is critical for establishing the size, density, and composition of transiting planets.  A precise measurement of the transit duration can constrain the stellar density (and hence the stellar size) for planets on circular orbits (e.g., Torres et al.\ 2008) and/or help to characterize the orbital eccentricity (e.g., Ford et al. 2008; Bakos et al.\ 2009).  Thus, the near degeneracy between the impact parameter and stellar limb darkening model will affect all $Kepler$ observations (and early CoRoT discoveries) that use a single, broad, white broadband filter, which can be severely affected by stellar limb darkening.  Fortunately, ground-based follow-up photometry is routinely performed at near-infrared (NIR) wavelengths which are only minimally impacted by limb darkening (e.g., observing with a Sloan $i'$ or $z'$ filter; see Holman et al. 2006, for example, as well as \S\ref{SecGround} in this paper).  Thus, NIR photometry (or space-based IR photometry; e.g., Nutzman et al. 2009) can enable a more precise measurement of the impact parameter, planet-star radius ratio, and transit duration (as shown in \S\ref{SecResults}). Ground-based follow-up photometry may offer additional advantages for very high cadence observations to search for transit timing variations (Agol et al.\ 2005; Holman \& Murray 2005; Ford \& Holman 2007) and/or changes in the transit duration or shape that could be caused due to moons, rings, or tidal bulge (Kipping 2009; Ragozzine \& Wolf 2009).

Our investigation of the potential contributions of ground-based transit follow-up observations is further motivated by the improving precision of ground-based photometry for transit follow-up applications.  Once the host star and transit times are known, nearly routine ground-based transit follow-up observations can provide photometry with a precision of $\sim$1 millimagnitude (mmag) at a rate of $\sim$1 observation per minute for 11th and 12th magnitude stars, even with a relatively small observatory [e.g., 1.2m Fred L. Whipple Observatory (FLWO); Holman et al.\ 2006; Winn et al.\ 2007].  Recent observations with larger telescopes have achieved even higher photometric precisions for favorable planet-host stars, e.g., 0.39 mmag/min in $z'$-band for a $V$=12 star with the 2.2m University of Hawaii (UH) telescope (rescaled from the quoted precision for a $V$=12.7 star; Johnson et al.\ 2009) and 0.315 mmag/min in $z$-band for a $V$=12 star with the 6.5m Magellan telescope (rescaled from the quoted precision for a $V$=12.5 star; Winn et al. 2009).  It is particularly noteworthy that differential photometry largely corrected for atmospheric effects, as demonstrated by high-precision ground-based photometry that has reached a precision very close to that of the theoretical limit (e.g., Winn et al.\ 2009).  This raises the possibility that modern instrumentation at a large observatory might soon be able to routinely achieve a photometric precision comparable to that of CoRoT or $Kepler$ (Gillon et al.\ 2008).

With the $Kepler$ mission's recent launch (March 2009)\footnote[2]{http://kepler.nasa.gov/} and CoRoT's discovery of $\sim$6 new transiting planets to date\footnote[3]{As of July 2009; http://exoplanet.eu/}, the future looks promising for finding many transiting planets with a range of sizes and orbital periods.  It will be necessary to use all of the tools at hand in order to describe the overall population of extrasolar planets in terms of both orbital and physical parameters.  Combining ground-based photometric observations with space-based transiting planet searches is a valuable tool, as astronomers push to constrain properties of potentially habitable Earth-like planets. 

This paper explores the potential benefits of adding ground-based NIR photometry to higher-precision transit photometry measurements in white light soon to be made by  $Kepler$.  In \S\ref{SecSims} we describe how we generate simulated transit light curves (LCs), and we describe our analysis procedure for fitting a transit model to each light curve in \S\ref{SecAnalysis}.  In \S\ref{SecResults}, we report the results in terms of the precision with which physical and orbital parameters (e.g., radius ratio, planet velocity projected onto the plane of the sky) can be measured.  Section \ref{SecKepler} (\ref{SecGround}) describes the results from $Kepler$-like (ground-based) observations alone.  We report results for combining $Kepler$-like and ground-based observations in \S\ref{SecGS}.  In \S\ref{SecIncreased} and \ref{SecFaint}, we consider the potential of higher-precision ground-based observations and the impact of a fainter target star.  Finally, we summarize our results and discuss the implications for future ground-based follow-up campaigns in \S\ref{SecDiscuss}.

\section{Methods}
\label{SecMethods}

In \S2.1 we describe our model for generating simulated transit light curves, intended to mimic $Kepler$-like photometry and ground-based follow-up observations (in the NIR; see also \S\ref{SecGround}).  In \S2.2, we discuss our analysis of the light curves, including the fitting process and how we quantify measurement uncertainties.

\subsection{Simulations of Transiting Planet Light Curves}
\label{SecSims}

We generated mock transit light curves for $Kepler$-like and ground-based observations, using the code from \citet{mandel2002} to incorporate realistic limb darkening.  The projected separation from the center of the planet to the center of the star ($z$) is calculated as 
\be
z=\sqrt{b^2+v^2(t-t_0)^2},
\ee
where $b$ is the impact parameter in units of stellar radii, $v$ is the planet velocity projected onto the plane of the sky, and $t_0$ is the time of mid-transit.  In this approximation, the radial star-planet separation is assumed to be constant over the transit duration.  Since the transit duration is much shorter than the orbital period, this is an excellent approximation for all but highly eccentric orbits.  Here, we define the total transit duration ($D$) as
\be
D=\frac{2\sqrt{(1+p)^2-b^2}}{v},
\ee
where $p$ is the planet-star radius ratio, defined as $p \equiv R_p/R_{\star}$.  We use the \citet{claret2000} quadratic limb darkening law, which has the following form, 
\be
I(r)=\frac{1-u_1(1-\sqrt{1-(r/R_{\star})^2})-u_2(1-\sqrt{1-(r/R_{\star})^2})^2} {(1-u_1/3-u_2/6)/\pi},
\ee
as given in \citet{mandel2002}.  The variables $u_1$ and $u_2$ are the linear and quadratic limb darkening (LD) coefficients.  

For the purposes of simulating $Kepler$-like photometry, we consider a range of planet radii (from 1 to 12 $R_{\oplus}$) and orbital periods (from 1 day to 1 year) to generate the appropriate light curves.  For our primary analysis, we consider planets on circular orbits, solar-type stars with stellar radii of one solar radius, and a representative value of 0.5 for the impact parameter.  For the cases where we consider a near-central transit, we adopt an input impact parameter of 0.01.  Finally, we use limb darkening coefficients of 0.4382 and 0.2924 ($u_1$ and $u_2$) based on ATLAS models for a solar-type (G2V) star in the $V$-band \citep{claret2000}.  We considered the same specifications for simulating ground-based photometry, but different limb darkening coefficients were used (see \S\ref{SecGround}).  Note that we chose these specifications considering that the $Kepler$ mission is designed to search for Earth-sized planets in the habitable zone.  Given the mission lifetime and abundance of bright stars, most of the $Kepler$ target stars are expected to range from late F stars to early K stars.  The nominal visual magnitude range is 9 to 15.  The mission specifications were chosen such that $Kepler$ could detect an Earth-size planet in a one year orbit around a solar-type star with $V=12$.\footnote[4]{http://nspires.nasaprs.com/external/solicitations/summary.do?method=init\&solId={0B53C123-58C9-69F7-D74B-9E7A1A5DD8E1}\&path=open}  If small planets are common, then there is a chance that Earth-size planets could be found around the relatively modest number of even brighter stars in the $Kepler$ field (e.g., Rouan et al. 2009).\footnote[5]{http://www.colloquium.eu/congres/09COROT/docs/slides/03mardi/11h/d\_rouan/alancer.pdf}

Finally, when generating mock light curves for $Kepler$-like targets, we inject realistic measurement errors.  We assume uncorrelated Gaussian noise and add this to each observation in each light curve.  For space-based observations, we assume a photometric precision of 0.4 mmag for each 1 minute integration, based on the expected precision from $Kepler$ observations for a $V = 12$ star \citep{basri2005}.  Note that the typical integration time employed by $Kepler$ will be 30 minutes, but for planet candidates, a fast cadence mode (i.e., 1 minute integration) will be used (Borucki et al.\ 2009).  For ground-based observations, we consider multiple precisions: one relatively modest precision (1.7 mmag) which is routinely achievable with 1-meter class telescopes and for which a $1/\sqrt{N}$ scaling has been demonstrated on timescales of up to $\sim$30 minutes \citep{winn2007}, and two relatively high precisions of two and four times better than the low precision (i.e., 0.85 and 0.425 mmag) for which it is not known if errors will scale as $1/\sqrt{N}$.  We also assume uncorrelated Gaussian noise for all the ground-based precision cases.  See \S\ref{SecGround} for a more detailed discussion of the generation of mock light curves for ground-based observations conducted in the NIR.  Once the light curves were generated, we fit and analyzed them using the process described in the following section (\S\ref{SecAnalysis}).

\subsection{Analysis of Transiting Planet Light Curves}
\label{SecAnalysis}

In order to analyze our simulated light curves, we employed the Levenberg-Marquardt least-squares method, which identifies a ``best fit'' model for each simulated light curve by minimizing $\chi^2$.  We used the publicly available code $mpfitfun$\footnote[6]{The code is available via Craig Markwardt's website, http://www.physics.wisc.edu/$\sim$craigm/idl/idl.html} for the fitting process, which allowed us to vary each of the following parameters that define the light curves: $b$, $v$, $t_0$, $c_1$, $c_2$, and $p$.  Here, we have employed an alternative parameterization of limb darkening where $c_1 \equiv u_1+u_2$ and $c_2 \equiv u_1-u_2$.  In the fitting process, we restrict the values of the parameters for $b$, $c_1$, $c_2$, and $p$ to the intervals [0,0.99], [-1,1.5], [-1,1.5], and [0,1].  In principle, one could adopt fixed values for the limb darkening parameters based on other observations (stellar spectral type, temperature, metallicity, surface gravity) and stellar atmosphere models.  However, this would result in underestimating the uncertainties for the transit parameters and potentially introduce biases if the spectroscopy and/or models are not accurate (Southworth 2008).  For sufficiently large planets, we expect that the transit photometry will provide stronger constraints on the limb darkening parameters, so that there is little benefit to incorporating assumptions about limb darkening.  For the smallest planets ($\sim1$ $R_{\oplus}$), measuring even basic features of the transit light curve will be a challenge, so it may be advantageous to constrain the limb darkening parameters based on stellar properties and atmosphere models, at least for stars where the models have been well tested.  In order to test the sensitivity of our conclusions to assumptions about the limb darkening coefficients, we reanalyzed a subset of our simulated observations three ways:  1) with both limb darkening parameters fixed at their true values, 2) the quadratic coefficient fixed and the linear coefficient allowed to vary, and 3) both limb darkening coefficients allowed to vary during the fitting process.  We find that the same basic conclusions can be made for each case despite any systematic shifts in the numerical results.  Therefore, we treat both the linear and quadratic limb darkening coefficients as free parameters during the fitting of our transit light curves.  See \S\ref{SecMultiGS} for a further discussion of how fixing or fitting various limb darkening terms affects the conclusions of this paper.

Once a ``best fit'' was found, the fitting routine returned two estimates of the uncertainties: 1) the formal $1\sigma$ error estimates obtained from the covariance matrix and 2) the bootstrap error estimates.  Here, we have defined the bootstrap errors as the median absolute deviations (MAD; median $|\delta v_{fit,true}|$) between the best-fit values and the true parameter value used to generate the given light curve.  We report median absolute deviation rather than standard deviation because in some cases a small fraction of the data sets result in light curve parameters that differ from the input parameters by significantly more than is typical.  The median absolute deviation is a more robust statistic that characterizes the typical deviation between the input and best-fit values.  Also note that for a normal distribution, the $1\sigma$ error is $\approx1.48 \times $MAD. 

We focus on how accurately the planet velocity and planet-star radius ratio could be measured when limb darkening is taken into account, and we report fractional error estimates for these parameters (i.e. the error estimates normalized by the true value for each of these parameters).  For example, we use the normalized $1\sigma$ errors, $\sigma_v/v_{true}$ (hereafter simply $\sigma_v/v$) to quantify the quality of the velocity measurements.  This provides a direct method of estimating the measurement uncertainties in the transit duration.  We use similar definitions to quantify the radius ratio estimates.  Note that in order to accurately compare uncertainty estimates, we used the median absolute deviation taken from the results of fifty sets of simulated light curves for each set of input parameters.  Unless otherwise specified, we report error estimates based on the covariance matrix.  This choice is based on an initial comparison of the error estimates.  While the $1\sigma$ and bootstrap errors are otherwise roughly comparable, we find a large discrepancy between our two error estimates when analyzing light curves due to planets with small radii ($R_p \lesssim$ 2 $R_{\oplus}$).  This is because for general non-linear models, these two uncertainty estimates are not always comparable, even for Gaussian noise (e.g., Ford 2005).  While the bootstrap analysis quantifies how much the best-fit model parameters change due to different realizations of the random noise, the covariance matrix analysis describes how rapidly $\chi^2$ increases as one varies the model parameters from their best fit values.  Therefore, the bootstrap analysis can underestimate the uncertainties when one of the parameters is pushed to the limit of its allowed range (e.g., impact parameter of zero for central transits).  Note that these conditions hold even if the limb darkening parameters are held fixed.  Accordingly, we believe the $1\sigma$ errors from the covariance matrix are most representative of the measurement precision.

\section{Results}
\label{SecResults}

In this section we present results for the expected measurement precision for the planet velocity and the planet-star radius ratio based on the analysis of simulated photometric light curves.  Several different cases are considered.  The precisions obtained from $Kepler$-like photometry and ground-based photometry are discussed in \S\ref{SecKepler} and \S\ref{SecGround}.  The fractional accuracy obtained as a result of combining space-based light curves with either one (multiple) ground-based light curve(s) is presented in \S\ref{SecOneGS} (\S\ref{SecMultiGS}).  The final two sections discuss the benefit of having increased ground-based photometric precision (\S\ref{SecIncreased}) and the measurement accuracies obtained when fainter stars (e.g., $V=14$) are observed (\S\ref{SecFaint}).

\subsection{$Kepler$ Photometry}
\label{SecKepler}
\subsubsection{Fractional Accuracy Curves}
\label{SecKeplerA}

The $Kepler$ mission is expected to detect hundreds of transiting planets that are predicted to range from sub-Earth-size to giant planets, with orbital periods from days to a year \citep{basri2005}.  Here, we estimate the measurement precision for the planet velocity and the planet-star radius ratio assuming $Kepler$-like photometry for a few fiducial cases.  In Table \ref{tab_one} we list the precisions obtained from our analysis of the simulated transit light curves of a Neptune-size planet ($R_{Nep}\approx$ 4 $R_{\oplus}$) and of a Jupiter-size planet ($R_{Jup}\approx$ 11 $R_{\oplus}$) orbiting a solar-type star with periods of three days (short-period) or one year (long-period).  We assume observational parameters as described in \S\ref{SecSims} for a non-central transit ($b = 0.5$).  Despite the effect of limb darkening, we find that the accuracy with which the velocity and radius ratio can be measured is better than 10\% in all cases considered (for solar-like stars).  The best-case scenario is the detection of a short-period Jupiter, with an accuracy of about 0.1\% for velocity and 0.08\% for the radius ratio.  In general, we find that planets with shorter periods allow for the best constraints on light curve parameters due to the high number of transit light curves available for analysis (and thus the larger amount of in-transit data available).  We assume an observing span of $\sim$2.5 years, so that the number of light curves acquired (and included in the fitting process) for planets with 3 day, 30 day, and 1 year orbital periods is 304, 30, and 3.  Since an individual transit duration scales as $\sim P^{1/3}$, and the rate of transits scales as $\sim P^{-1}$, the integrated time in transit scales as $\sim P^{-2/3}$, so the planet-star radius ratio of short-period planets will be generally be better constrained than otherwise identical long-period planets if only space-based observations are available (for similar host stars).  The same scalings apply for the time spent in ingress/egress, so parameters such as the planet velocity and eccentricity will also be best constrained for short-period planets (until time sampling becomes inadequate).  While it becomes increasingly difficult to obtain precise measurements for planets with longer orbital periods and smaller radii, adding ground-based observations can improve upon this (see \S\ref{SecGS}).

Besides the fiducial cases, it is also useful to examine the precision over larger ranges of planet sizes and orbital periods.  In Figure \ref{FigKepler} we show estimated fractional accuracies for the planet velocity and radius ratio from $Kepler$-like photometry of a non-central transit as functions of planet radius and orbital period.  Recall that the total number of light curves used in the fitting process is a function of the orbital period and the assumed lifetime of the $Kepler$ mission.  In all cases the radius ratio is determined with higher precision than the planet velocity.  We find that the size of an Earth-size planet can be measured to about 20\% accuracy at any orbital period, but the velocity is very poorly constrained with approximately 80\% accuracy.  While an accuracy of 80\% is effectively a non-detection, even a weak constraint on the velocity of small transiting planets will be useful for characterizing the overall eccentricity distribution (Ford et al. 2008).  The velocity and radius ratio of Neptune-size planets can be measured to accuracies of about 2\% and 0.5\% for short periods.  At longer periods, the accuracy of measurements for Neptune-size planets is approximately 10\% and 3\% for the velocity and radius ratio.  

For a near-central transit ($b = 0.01$; not shown), constraints on the size of Earth-size planets are more sensitive to orbital period, with measurement accuracies of approximately 1\% at short periods and 5\% at longer periods.  For Neptune-size planets, the velocity and radius ratio are both constrained with accuracies of about 0.08\% at short periods.  At longer periods, the velocity and radius ratio are determined with uncertainties of about 0.5\%.  These results indicate that photometric observations of a central transit offer better constraints on parameters than observations of a non-central transit.  Possible reasons why observations of a central transit offer more accurate measurements include there being smaller effects from limb darkening (i.e. the effect of limb darkening is more easily distinguished from ingress/egress) and more observations in transit (due to a longer transit duration).  Despite the higher quality of measurements obtained from observing central transits, this paper primarily focuses on the more representative case of observing non-central transits.

\subsubsection{Analytic Fits}
\label{SecKeplerB}

\begin{figure}
\plotone{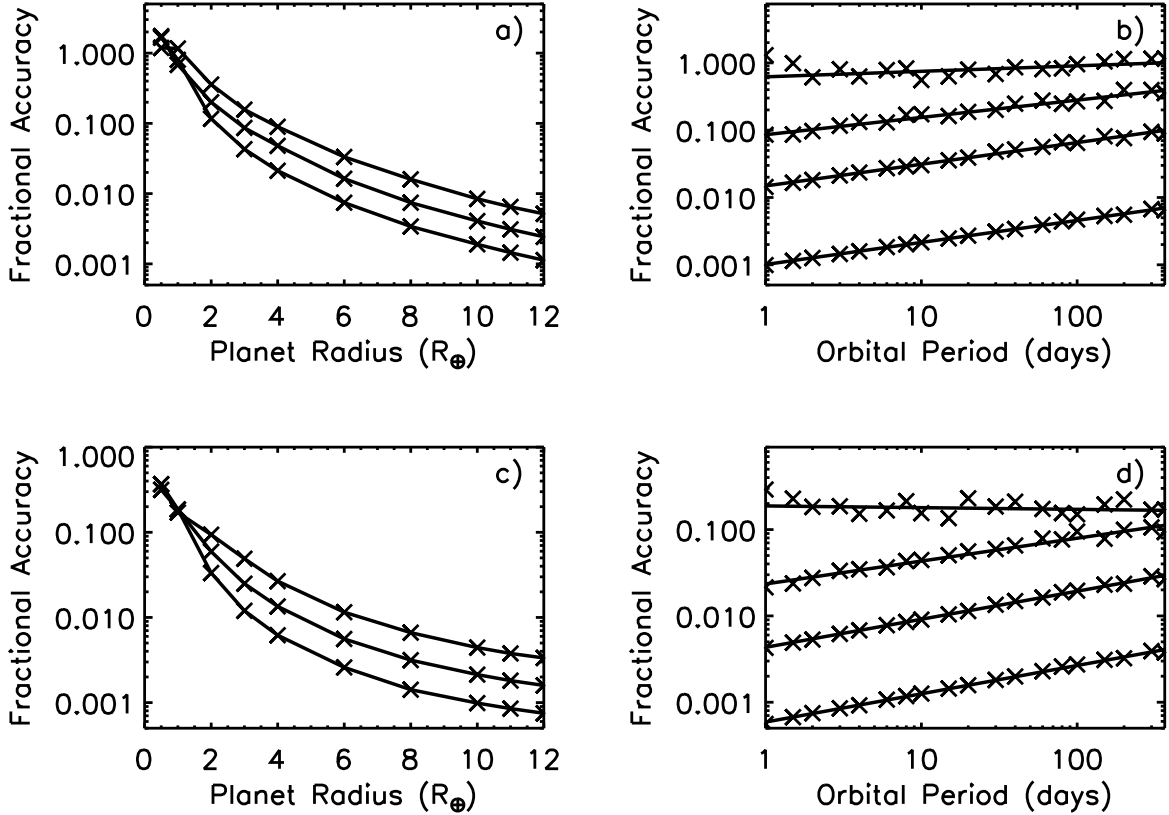} % (paper_figs12.pro)
\caption{Estimated fractional accuracies for velocity (top; a,b) and radius ratio (bottom; c,d) based on simulated $Kepler$-like photometry for a non-central transit ($b = 0.5$).  Left-hand panels (a,c) show accuracy (crosses) for orbital periods of 1 year, 30 days, and 3 days going top to bottom, and lines are shown to guide the eye.  The right column shows results (crosses) for different size planets, going from 1 $R_{\oplus}$, 2 $R_{\oplus}$, 4 $R_{\oplus}$, and 11 $R_{\oplus}$ top to bottom, and the solid lines represent analytic fits for each planet (results of the fits are given in Table \ref{tab_three}).  The total number of light curves used in the analysis are 3, 30, and 304 for 1 year, 30 days, and 3 days in panels (a) and (c) and vary based on orbital period and the assumed lifetime of the $Kepler$ mission in panels (b) and (d).
\label{FigKepler}}
\end{figure}

For a semi-analytic method of estimating the expected fractional uncertainties based on $Kepler$-like observations, we fit power laws to our results for the accuracies of both planet velocity and radius ratio.  The accuracies were fit as a function of planet radius (specifically for radii ranging from 4 $R_{\oplus}$ to 12 $R_{\oplus}$) for fiducial periods of three days, thirty days, and one year.  We also present the accuracies as a function of orbital period for planets with radii of 1 $R_{\oplus}$, 2 $R_{\oplus}$, 4 $R_{\oplus}$, and 11 $R_{\oplus}$.  Results are given in Tables \ref{tab_two} and \ref{tab_three}, and the analytic fits as a function of orbital period for the non-central transit case are shown in Figure \ref{FigKepler}.  The trend in the accuracy as a function of planet radius is steep for non-central transits, with power-law exponents of about -2.6 to -2.7 and -1.9 for the velocity and radius ratio accuracies.  Most of the accuracies given as a function of orbital period have a power-law exponent near the expected value of 1/3, with the exception of Earth-size planets.  The lack of a clear improvement in measurement precision with decreasing orbital period further emphasizes the need for additional data to be used in conjunction with space-based observations of small planets (\S\ref{SecGS}). 

We find that our power-law fits are most comparable to those found by \citet{carter2008} and \citet{ford2008} if we assume a central transit and hold the limb darkening parameters fixed.  In that case, the power-laws go as $R_p^{-3/2}$ and $P^{1/3}$ for the transit duration/planet velocity accuracies.  In general, we find that the power-laws from our non-central transit models are significantly steeper than for central transits, even when we keep the limb darkening coefficients fixed.  This implies that limb darkening has a more significant effect on measurements of orbital parameters for non-central transits.  It should also be noted that when limb darkening is fitted, a single power-law cannot be fitted over the entire range of planet radii shown in Figure \ref{FigKepler}, which is why we fit a power law only over the range of 4 to 12 $R_{\oplus}$.  This is most likely because limb darkening cannot be well-characterized from $V$-band observational data alone for smaller planets ($R_p \lesssim$ 4 $R_{\oplus}$).  All of this should be kept in mind when planning follow-up observations of any transiting system whose light curve is strongly influenced by limb darkening.

\subsection{Ground-Based Observations}
\label{SecGround}

Before combining ground-based observations with space-based observations, we considered which ground-based observation band would provide the most accurate measurements by comparing the accuracies obtained from ground-based photometry in the $r'$-, $i'$-, and $z'$- bands.  We considered the same photometric precision of 1.7 mmag each minute for each band based on the typical precision obtained by, e.g, Charbonneau et al. (2007), \citet{winn2007}, and Sozzetti et al. (2009) with the 1.2m telescope at FLWO.  In order to simulate a generic data set, we assume the time sampling of 1 minute takes into account the exposure time, readout time, setup time, and any other dead time between observations.  Recall that \citet{winn2007} demonstrate that their photometric errors decrease as $1/\sqrt{N}$ when binned over timescales of up to $\sim$30 minutes, justifying our assumption of uncorrelated noise.  We assumed a quadratic limb darkening law, with coefficients for a solar-type star determined from ATLAS models \citep{claret2004}.  The linear and quadratic coefficients ($u_1,u_2$) used are (0.433, 0.30215), (0.3183, 0.3375), and (0.22785, 0.3573) for the $r'$-, $i'$-, and $z'$- bands.  Fractional accuracies for the fiducial cases of planets of 4 $R_{\oplus}$ and 11 $R_{\oplus}$ and with orbital periods of three days and one year are presented in Table \ref{tab_four}.  The differences in accuracies are primarily a result of the different limb darkening coefficients used for each filter.  Assuming a common measurement precision, we find that the $z'$ filter typically offers the best precision, largely because limb darkening is less significant at longer wavelengths (also see P\'al 2008).  It should be noted that for relatively noisy data, the $i'$ and $z'$ filters yield comparable measurement accuracies.  Also, recent observations by \citet{winn2008} achieved a better photometric precision of 0.77 mmag/min for a $V$=12 star (rescaled from the quoted precision for a $V$=12.4 star) in the $i$-band with the FLWO 1.2m telescope.  If a factor of two improvement in precision were routinely obtained, then the $i$-band would be equally useful compared to the $z'$-band.  Other recent observations in the $z'$-band resulted in an even better photometric precision of 0.39 mmag/min for a $V$=12 star (rescaled from the quoted precision for a $V$=12.7 star) using the 2.2m UH telescope \citep{johnson2009}, which indicates that the level of ground-based photometric precision achieved is improving even as we write this paper.  (Note that while this technique results in errors that scale as $1/\sqrt{N}$ over $\sim$20 minute timescales, it has yet to be demonstrated that observations with this level of noise will consistently follow that scale over longer timescales.)  The key point is that limb darkening plays a large role in reducing the measurement precisions for a planet's size and orbital parameters, making it especially important to minimize its effects.  Therefore, we conclude that it is highly desirable to obtain some NIR observations (e.g., using an $i'$ or $z'$ filter) when doing transit photometry from the ground.  In the following sections, we use simulated ground-based observations conducted in the $z'$-band to determine the benefits of adding them to $Kepler$-like observations.

\subsection{Combining Ground- and Space-Based Observations}
\label{SecGS}
In this section we present results obtained from analyzing a combination of simulated ground-based and space-based observations to determine the planet velocity and radius ratio.  We specifically discuss the improvements in measurement accuracy obtained when either one or several simulated ground-based light curves (as described in \S\ref{SecGround}) are combined with the $Kepler$-like photometric observations described in \S\ref{SecKepler}.  
\subsubsection{Addition of One Ground-Based Light Curve}
\label{SecOneGS}
\begin{figure}
\plotone{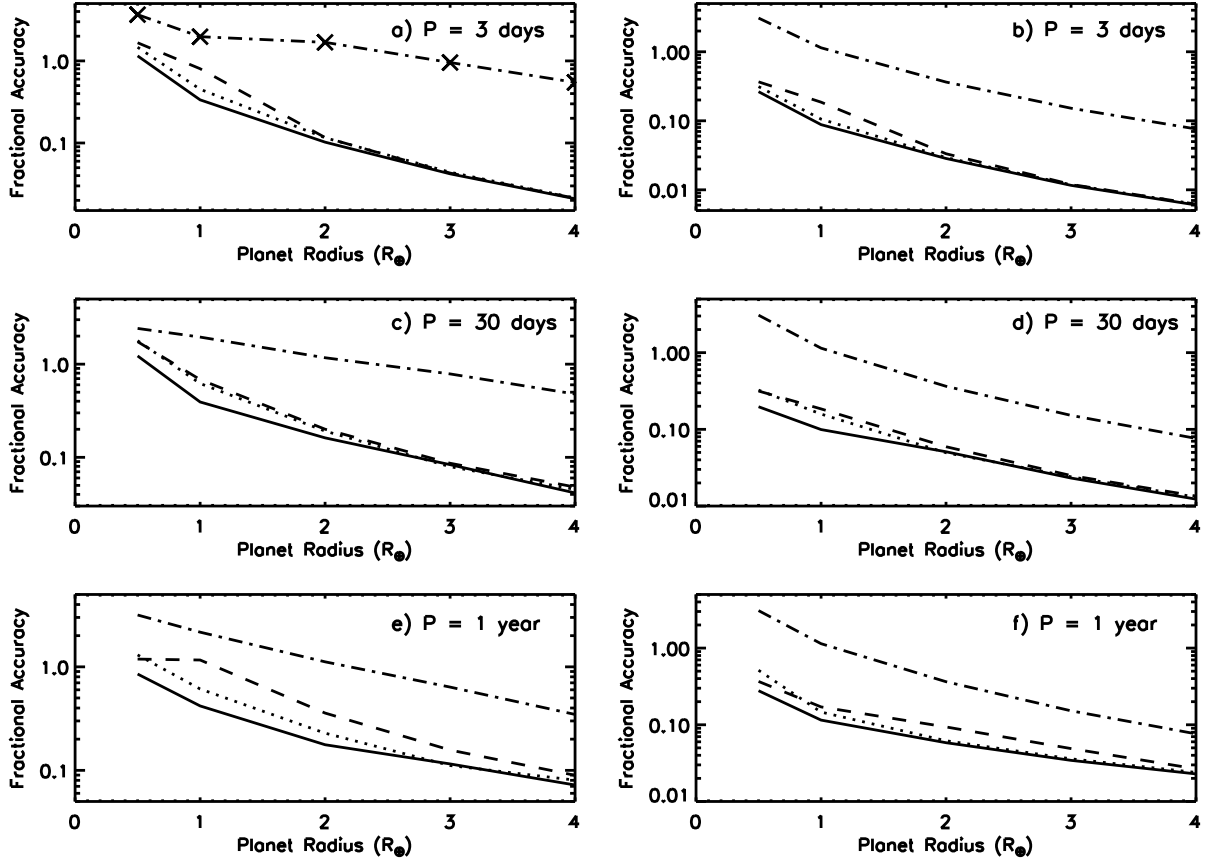} % (paper_figs34.pro)
\caption{Estimated fractional accuracies of the planet velocity and radius ratio as a function of planet radius for various observational programs.  The left column shows velocity accuracies; the right column shows radius ratio accuracies.  Curves are based on observations from the ground only ($z'$ filter; dot-dashed curve), space only ($Kepler$; dashed curve), space plus one ground (dotted curve), and space plus multiple ground (solid curve).  The number of multiple ground curves added varies based on the orbital period.  Sixteen, eight, and two ground-based LCs were added to space-based LCs for orbital periods of three days, thirty days, and one year.  The x-coordinate of the crosses in panel (a) indicate the actual radii used in the simulations. 
\label{FigAllRad}}
\end{figure}

\begin{figure}
\plotone{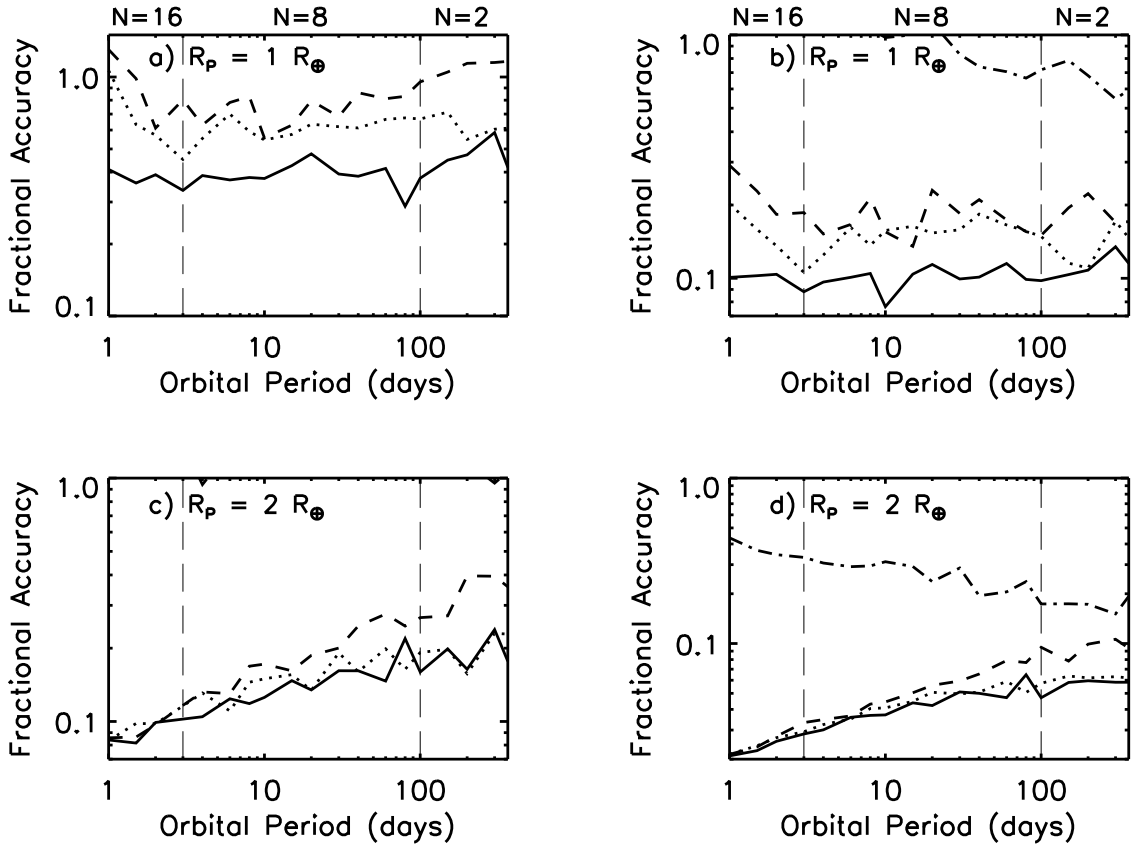} % (paper_figs34.pro)
\caption{Estimated fractional accuracies of the velocity and radius ratio versus orbital period for 1 $R_{\oplus}$ and 2 $R_{\oplus}$ planets.  The left column contains velocity accuracies; the right column has radius ratio accuracies.  The linestyles are the same as in Fig. \ref{FigAllRad}.  Note that in panels (a), (b), and (c), the dot-dashed curve representing ground-based accuracy is either not shown or only partially shown due to the fractional accuracy being greater than 1.  Different numbers of ground-based LCs were added to the standard $Kepler$-like LCs according to orbital period.  The number of ground-based LCs added are given and the period intervals are marked as long-dashed vertical lines in all panels.
\label{FigAllPer}}
\end{figure}

The fractional accuracy curves for ground-based observations only, space-based observations only, and space plus either one or multiple ground-based light curves (for a non-central transit) are shown in Figures \ref{FigAllRad} and \ref{FigAllPer}.  The estimated fractional accuracies of planet velocity (left columns) and radius ratio (right columns) as a function of both planet radius (Fig. \ref{FigAllRad}) and orbital period (Fig. \ref{FigAllPer}) are shown.  While the individual ground-based curves (dot-dashed) are clearly less accurate than the space-based curves (dashed) in all cases, adding just one ground-based LC to the space-based LCs does in fact increase the measurement precision (e.g., dotted curves).  As emphasized in Figures \ref{FigAllRad} and \ref{FigAllPer}, the results are dependent on both planet size and orbital period.  First, note that only small planets ($R_p \lesssim$ 4 $R_{\oplus}$) gain significant measurement improvements from the simulated follow-up ground-based observations.  This implies that ground-based photometric follow-up of the confirmed primary transits of Jupiter-size planets detected around relatively bright stars ($V \sim 12$) by $Kepler$ should be a low priority.  (Note that it is beyond the scope of this paper to investigate the benefit of ground-based observations for confirming the planetary nature of candidates or other types of observations, such as occultation.)  Second, the space-only and space-plus-one-ground accuracy curves are barely distinguishable at short periods, but at long periods, the difference between the two curves is significant for planetary radii of up to approximately 3 $R_{\oplus}$.  Figure \ref{FigAllPer} specifically illustrates the change in accuracy as a function of orbital period for planets of 1 and 2 $R_{\oplus}$.  As discussed in \S\ref{SecKepler}, $Kepler$-like photometry only allowed the velocity of an Earth-size planet to be determined with a very weak constraint (about 80\% accuracy), and the radius ratio to be determined to an accuracy of about 20\%.  With the addition of one ground-based LC, the accuracy with which the velocity and radius ratio are measured becomes about 70\% and between 15\% and 20\% at long periods.  If $Kepler$ detects many short-period planets then even loose constraints on the planet velocity can be useful for constraining the eccentricity distribution \citep{ford2008}.  Despite the fact that the ground-based LCs have about four times worse precision than the simulated $Kepler$ observations, combining ground and space data helps constrain the impact and limb darkening parameters in addition to improving the measurements of a long-period planet's parameters because the total time of in-transit observations from the ground for a long-period planet is a larger fraction of the total in-transit observations available for that same planet over the lifetime of the $Kepler$ mission.  In essence, adding 1 ground-based LC to only 3 space-based LCs (for a long-period planet) results in a larger overall improvement compared to adding 1 ground-based LC to 304 space-based LCs (for a short-period planet).

For the case of a central transit (not shown), we find that adding one ground-based observation to $Kepler$ observations barely increases the accuracy with which the radius ratio is measured.  For velocity measurements, there is a slight benefit obtained for smaller planets ($R_p \lesssim$ 2 $R_{\oplus}$).  Therefore, the primary cases that benefit from a combination of at least one ground-based observation with space-based ones are non-central transits of small planets ($R_p \lesssim$ 4 $R_{\oplus}$) that have longer orbital periods.  In all cases, the radius ratio is measured more accurately than the planet velocity, but the velocity measurements have a larger absolute improvement upon fitting a ground-based LC in combination with space-based observations.  Finally, it is important to realize that the transit duration of a 1 $R_{\oplus}$ planet in a 1-year orbit is about 11.5 hours (for a non-central transit), which means it is not possible to observe an entire transit from one location.  Therefore, collaborations between observers in different parts of the world are key to obtaining complete LCs for long-period planets.  Given the substantial improvement in measurement accuracy from the addition of one ground-based LC alone, observing a transit of a long-period planet is extremely valuable and worth the observing time.  

\subsubsection{Addition of Multiple Ground-Based Light Curves}
\label{SecMultiGS}
The benefits of adding multiple ground-based light curves to space-based observations are illustrated in Figures \ref{FigAllRad} and \ref{FigAllPer} as the solid curves.  Note that different numbers of ground-based LCs were fitted in combination with space-based LCs depending on the orbital period (16, 8, and 2 ground-based LCs were fitted for periods of 3 days, 30 days, and 1 year).  We selected the number of LCs to use in each analysis based on the number of ground-based observations that could realistically be completed over the lifetime of the $Kepler$ mission.  For non-central transits, we find that adding multiple ground-based LCs of short-period planets with radii less than about 3 $R_{\oplus}$ results in a larger overall improvement in both velocity and the radius ratio when compared to adding only one ground-based LC (see \S\ref{SecOneGS}).  For longer periods, there is still a slight improvement from adding multiple ground-based LCs, but adding even one suffices to increase the measurement accuracy.  There is still no significant benefit seen for planets with radii larger than $\sim$4 $R_{\oplus}$.

For central transits of small planets at longer periods (not shown), there is no significant improvement from adding multiple ground-based LCs, just as there was none seen upon adding just one ground-based LC.  However, for smaller planets ($R_p \lesssim$ 2 $R_{\oplus}$) at short periods, there is a noticeable improvement in the measurement accuracy for the planet velocity when multiple ground-based LCs are fitted.  For example, fitting only space-based observations of an Earth-size planet resulted in an accuracy of about 2\%.  When multiple ground-based LCs are fit in conjunction with the space-based simulations, the measurement accuracy improves to about 0.8\%.  There is still no significant improvement seen in measuring the radius ratio for small planets with short periods.    

In summary, measurements of the planet velocity and the radius ratio from non-central transits of small planets with short orbital periods benefit most from combining multiple ground-based observations with space-based observations.  At long periods, adding just one high-quality ground-based LC provides the majority of the benefit.  For central transits of small planets, the only improvement that we find is for the velocity measurement and solely for multiple ground-based observations of short-period planets. Note that the general conclusions are supported by simulations using each of the three different limb darkening scenarios, as discussed in \S\ref{SecAnalysis}.  In the simulations with both limb darkening parameters fixed at their true values, we find that the benefit of adding ground-based observations is substantially reduced and only significant for small, long-period planets.  We caution that assuming a fixed value for both limb darkening parameters can result in significantly underestimating the uncertainties in other transit parameters (Southworth 2008).  Our simulations in which only the quadratic limb darkening parameter was held fixed (following the recommendation of Southworth 2008) show an improvement in the measurement precision for planets with radii $\le~2-4$ $R_{\oplus}$ (for orbital periods of 3 days to 1 year).  There is still a similar benefit to incorporating ground-based observations, as when fitting both limb darkening coefficients.  For small long-period ($\sim$1 year) planets, even a single ground-based transit observation may be sufficient to realize most of this benefit, while several transit observations may be necessary for small short-period planets ($\sim$3 days).  Based on these results, we further conclude that our approach of fitting both limb darkening parameters is reasonable.

\subsection{Increased Ground-Based Precision}
\label{SecIncreased}
As discussed in \S\ref{SecGround}, recent observations by \citet{winn2008} and \citet{johnson2009} have obtained a level of photometric precision down to 0.77 mmag/min in the $i$-band and 0.39 mmag/min in the $z'$-band (each rescaled from their respective quoted precisions).  These are factors of over two and four improvement compared to our assumed precision of 1.7 mmag/min in the $z'$-band.  Considering that ground-based photometric precisions to these levels have already been obtained, it is plausible to assume that ground-based observations may routinely achieve such a level of precision in the near future.  Therefore, we consider the improvement in the accuracy of estimating the planet velocity based on increased ground-based photometric precisions of two (0.85 mmag/min) and four (0.425 mmag/min) times our assumed routine precision of 1.7 mmag/min.  The improvement in measuring the radius ratio has not been considered here, as it is already quite well constrained.

\begin{figure}
\plotone{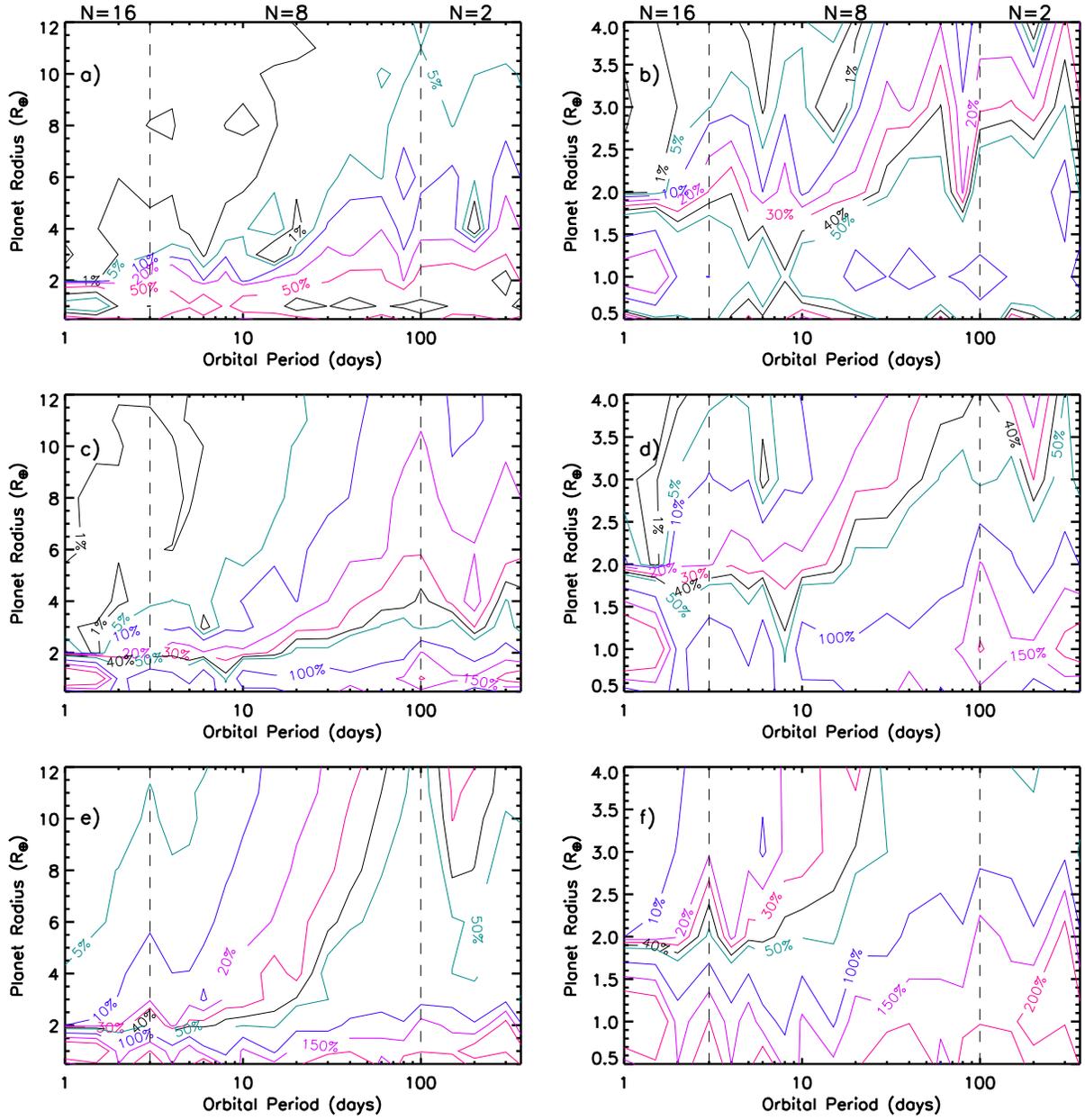} % contourx.eps (plot_rp2.pro)
\caption{Fractional benefit of increased ground-based precision for measuring planet velocity.  The benefit is defined as the accuracy obtained from space observations only divided by the accuracy from space plus multiple ground observations, minus unity.  Here we assume ground-based photometric precisions of 1.7 mmag/min (a,b), 0.85 mmag/min (c,d), and 0.425 mmag/min (e,f).  The number of ground-based light curves added varies with orbital period as indicated above panels (a) and (b) and with vertical dashed lines.  Each contour shows the percentage increase in the accuracy.  All panels show the same contours, but with the righthand panels showing a smaller range of planet radii.
\label{Fig2xContour}}
\end{figure}

The fractional benefit for measuring the planet velocity with increased ground-based precision is illustrated in Figure \ref{Fig2xContour}.  Here, we consider the accuracy obtained from only space observations divided by the accuracy obtained from space plus multiple ground observations (for a non-central transit).  We define the benefit to be that ratio minus unity.  The contours in Figure \ref{Fig2xContour} represent the percentage increase in the accuracy, based on the fractional benefit defined above.  As an example, the 1\% contour defines the region in which the velocity has been measured with 1\% more accuracy from including multiple ground-based observations in the fitting process.  The improvement in the velocity measurement accuracy is shown over a range of orbital periods and planet radii for photometric precisions of 1.7 mmag (Fig. \ref{Fig2xContour}$a,b$), 0.85 mmag (Fig. \ref{Fig2xContour}$c,d$), and 0.425 mmag (Fig. \ref{Fig2xContour}$e,f$).  The top panels of Figure \ref{Fig2xContour} ($a,b$) confirm that measurements of small planets (i.e. with radii less than 4 $R_{\oplus}$) benefit by as much as 100\% (i.e., a factor of two improvement) from the addition of multiple ground-based LCs that have a photometric precision of 1.7 mmag/min.  At this precision, larger planets with any orbital period have measurement improvements of only about 1-5\%.  However, when we consider an increased precision of 0.85 mmag/min (Fig. \ref{Fig2xContour}$c,d$), we find that there is a large region containing intermediate to long-period planets with radii larger than 4 $R_{\oplus}$ where an improvement between 10-20\% can be obtained.  At this increased photometric precision, the measurement accuracy for small planets also increases, with a maximum improvement of over 150\%.  Finally, if we assume a further improved photometric precision of 0.425 mmag/min (Fig. \ref{Fig2xContour}$e,f$), we find that measurement improvements of up to 50\% can be obtained for larger planets with intermediate to long periods.  Small planet measurements continue to improve by as much as 200\%.  In summary, as the photometric precision of ground-based observations continues to improve, there is an increasingly large range of planets that will benefit.  The present-day routine photometric precision of ground-based observations is not overly useful for characterizing large planets, but increasing precisions will allow for better constraints on almost any size planet in any size orbit.  However, there will continue to be a minimal benefit from adding ground-based observations of larger planets with shorter orbital periods.  

\subsection{Observing Fainter Stars}
\label{SecFaint}
Because the $Kepler$ mission will observe stars as faint as 15th magnitude \citep{basri2005}, we find it useful to explore the improvement in the measurement accuracy for derived parameters obtained when observing transits of 14th magnitude stars (in the $V$-band).  We considered a photometric noise of 1.0 mmag for each 1 minute observation of a $V=14$ star, based on rescaling the expected $Kepler$ precision for a $V=12$ star.  The assumptions made in \S\ref{SecGround} for the ground-based precision are also implemented here.  We considered a (low) ground-based photometric precision of 1.7 mmag each minute, which is comparable to the precision obtained by, e.g., Charbonneau et al. (2007), \citet{winn2007}, and Sozzetti et al. (2009) for a $V=12$ star with a 1-meter class telescope.  Here, we assume that a precision of 1.7 mmag/min will be obtainable for a $V=14$ star with a 2-meter class telescope.  Considering the recent improvements made in observing transits of $V \approx 12$ stars (e.g., \citep{johnson2009}), we also implement a better ground-based photometric precision of twice the lower precision, or 0.85 mmag/min.  To justify this as a realistic choice, we assume that a medium to large-size telescope would be able to achieve this level of precision.  We also acknowledge the anonymous referee for pointing out the possibility that it may be relatively easy to reach the photon noise limit for a 14th magnitude star compared to a brighter star, as fainter stars should have many more comparison stars within the observational field of view.

\begin{figure}
\plotone{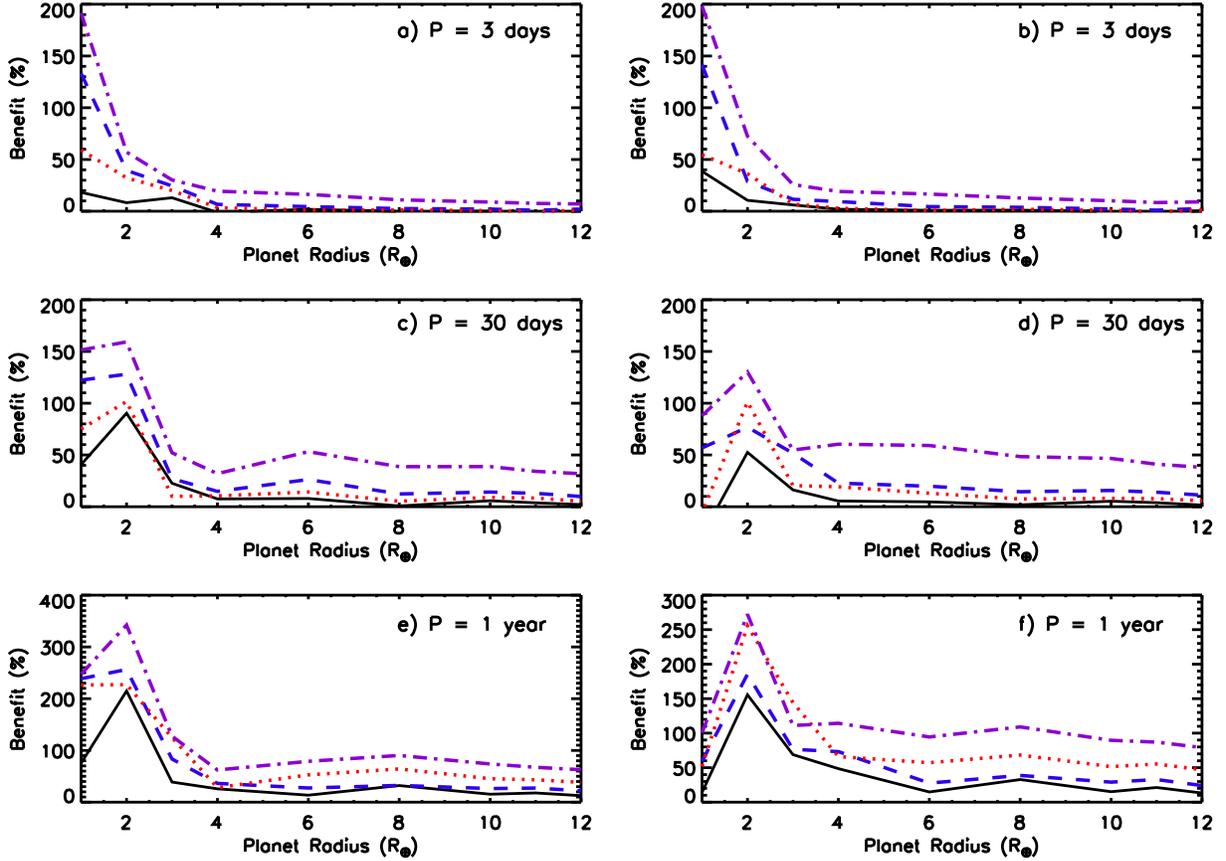} % (plot_3_5rad14new.pro)
\caption{Benefit of adding ground-based light curves to space-based observations for $V$=14 host stars. The benefit is defined as the accuracy obtained from space observations only divided by the accuracy from space plus either one or multiple ground observations, minus unity.  Planets with orbital periods of 3 days (a,b), 30 days (c,d), and 1 year (e,f) are shown.  The lefthand (righthand) column presents benefits for velocity (radius ratio) measurements.  Different observational scenarios are considered: adding one low-precision (1.7 mmag/min) ground-based LC (solid), multiple low-precision ground-based LCs (dashed), one high-precision (0.85 mmag/min) ground-based LC (dotted), and multiple high-precision ground-based LCs (dot-dashed).  We assume a precision of 1.0 mmag/min for $Kepler$-like observations of a $V=14$ star.
\label{FigV14}}
\end{figure}

Figure \ref{FigV14} compares the benefit of adding multiple ground-based LCs to space-based observations for $V=14$ stars with (non-central) transiting planets with 3 day, 30 day, and 1 year orbital periods.  We consider the addition of one low precision (1.7 mmag/min) ground-based LC (solid), multiple low-precision LCs (dashed), one high precision (0.85 mmag/min) LC (dotted), and multiple high-precision LCs (dot-dashed).  We find that incorporating one low-precision ground-based observation helps improve the accuracy of velocity and radius ratio measurements by as much as 200\% and 150\% for planets with $R_p \sim$ 2 $R_{\oplus}$ and long periods.  For both parameters, the benefit decreases with decreasing orbital period.  The benefits are greater when multiple low-precision ground-based LCs are added.  However, when only one high-precision ground-based LC is added for planets with long orbital periods, the benefit is greater than for adding multiple low-precision observations.  For short-period planets, the addition of multiple low-precision observations provides a larger benefit than one high-precision observation does.  Finally, the greatest benefits are obtained when multiple high-precision LCs are added, with an improvement in velocity and radius ratio measurements of about 350\% and 250\% for 2 $R_{\oplus}$ planets with long periods.  Overall, all cases benefit most from including ground-based observations for small planets, but the benefits are minimal for larger planets, similar to results in \S\ref{SecMultiGS}.

For near-central transits (not shown), we find that the overall benefits from adding ground-based observations are smaller than for the non-central transit cases, but are still appreciable.  Overall, this indicates that using ground-based observations in conjunction with space-based observations can be increasingly beneficial for constraining the velocity and radius ratio of small planets that transit stars fainter than 12th magnitude.

\section{Discussion}
\label{SecDiscuss}
Previous studies have provided analytic estimates of the precision of transit parameters for a single set of observations \citep{carter2008}.  In this paper, we investigate the precision of transit model parameters based on heterogeneous observational data sets.  In particular, we consider the case of high quality space-based photometry using a broadband optical filter and lower quality ground-based photometry in the NIR.  We have described how ground-based photometric observations of planets in transit can be useful for further constraining the planet velocity and planet-star radius ratio when they are combined with space-based observations obtained from missions like $Kepler$ and CoRoT.  We considered cases for both a near-central transit ($b=0.01$) and a non-central transit ($b=0.5$), where the effects of limb darkening are more significant.  Our results revealed that in all cases considered, the parameters measured for central transits are constrained better than identical parameters measured for non-central transits.  We find that the radius ratio is measured with higher accuracy than the planet velocity in accordance with analytic estimates that neglect limb darkening \citep{carter2008}.  Accordingly, the measurement of the planet velocity benefited most from adding ground-based photometric observations in conjunction with space-based ones, and the benefits were largest for planets with non-central transits and radii less than 4 $R_{\oplus}$.  We find a substantial increase in the measurement precision for the velocity of small non-centrally transiting planets with long orbital periods when at least one ground-based observation is added to space-based ones.  Multiple ground-based light curves added to space-based ones further improve the measurement precision for the velocity of planets and provide a substantial benefit for a wider range of planet sizes and orbital periods.  When higher ground-based precisions are considered, we find that any size planet in any orbital period can benefit from the addition of ground-based light curves.  Furthermore, even fainter stars (e.g., $V=14$) can benefit from the addition of relatively low precision ground-based observations.

Our results are important for those planning ground-based photometric follow-up of transiting planets that are initially discovered with space-based missions such as $Kepler$ and CoRoT.  For example, we have shown that routine ground-based follow-up can be particularly useful for planets with radii smaller than about 4 $R_{\oplus}$.  As ground-based observations push towards improved photometric precision, even the characterization of larger planets could benefit from the inclusion of NIR ground-based observations.  Thus, ground-based follow-up can improve the characterization of a large range of transiting planets, particularly for the planet velocity, transit duration, and the orbital eccentricity.  The constraints on the transit duration and eccentricity can further constrain planet formation models as well as help with the planning of follow-up observations of primary transit or occultation with warm-Spitzer, JWST, or ground-based observatories.

\acknowledgements We would like to thank Joshua Winn for several helpful comments.  We also gratefully acknowledge an anonymous referee, whose insightful comments and suggestions have greatly improved this paper.  This material is based upon work supported by the National Aeronautics and Space Administration under grant NNX08AR04G issued through the $Kepler$ Participating Scientist Program.  K.D.C.\ is supported by a University of Florida Alumni Fellowship.

\clearpage
              
\begin{deluxetable}{cccc} % Number of columns
\tabletypesize{\small}
\tablewidth{0pt}
\tablecaption{Fractional Accuracy of Planet Velocity and Size \label{tab_one}}
\tablehead{
\colhead{$R_p$} & \colhead{$P$} & \colhead{} & \colhead{} \\
\colhead{($R_{\oplus}$)} & \colhead{(days)} & \colhead{$\sigma_v/v$} & \colhead{$\sigma_p/p$}}
\startdata
$4$   & 3     & 0.0212  & 0.00619 \\
$4$   & 365   & 0.0933  & 0.0263 \\
$11$  & 3     & 0.00143 & 0.000824 \\
$11$  & 365   & 0.00651 & 0.00380 \\
\enddata
\tablenotetext{a}{All accuracies were determined based on simulated $Kepler$-like photometry of a non-central transit ($b = 0.5$) of a planet orbiting a 12th magnitude ($V$-band) solar-like (G2V) star.  Gaussian noise of 0.4 mmag was added to each 1 minute integration.  Limb darkening coefficients of $u_1 = 0.4382$ and $u_2 = 0.2924$ were used based on ATLAS models for a solar-like star \citep{claret2000} to generate the light curves, but we allowed them to vary when fitting the models.  The accuracies for the three day (1 year) orbital periods are based on fitting 304 (3) light curves.}
\end{deluxetable}
\clearpage

\begin{deluxetable}{ccccc}
\tabletypesize{\small}
\tablewidth{0pt}
\tablecaption{Fractional Accuracy as a Function of Planet Size \label{tab_two}}
\tablehead{
\multicolumn{1}{c}{$P$ (days)} & \multicolumn{1}{c}{$\sigma_v/v$} & \multicolumn{1}{c}{$\sigma_p/p$}}
\startdata
3    & 0.865$R_p^{-2.66}$ & 0.0841$R_p^{-1.92}$ \\
30   & 2.08$R_p^{-2.71}$  & 0.190$R_p^{-1.94}$  \\
365  & 3.46$R_p^{-2.61}$  & 0.357$R_p^{-1.90}$   \\
\enddata
\tablenotetext{a}{The analytical models were fitted for planets with radii ranging from 4 $R_{\oplus}$ to 12 $R_{\oplus}$ based on simulated $Kepler$-like photometry of a non-central transit (as described in Table \ref{tab_one}).  The radius of the planet, $R_p$, is in units of $R_{\oplus}$.}
\end{deluxetable}
\clearpage

\begin{deluxetable}{ccccc}
\tabletypesize{\small}
\tablewidth{0pt}
\tablecaption{Fractional Accuracy as a Function of Orbital Period\label{tab_three}}
\tablehead{
\multicolumn{1}{c}{$R_p$ ($R_{\oplus}$)} & \multicolumn{1}{c}{$\sigma_v/v$} & \multicolumn{1}{c}{$\sigma_p/p$}}
\startdata
1     & 0.620$P^{0.0817}$  & 0.188$P^{-0.0199}$ \\
2     & 0.0856$P^{0.257}$  & 0.0234$P^{0.265}$ \\
4     & 0.0148$P^{0.322}$  & 0.00432$P^{0.325}$ \\
11    & 0.00101$P^{0.327}$ & 0.000589$P^{0.328}$ \\
\enddata
\tablenotetext{a}{The analytical models were fitted for planets with periods from 1 day to 1 year based on simulated $Kepler$-like photometry of a non-central transit (as described in Table \ref{tab_one}).  The orbital period of the planet, $P$, is in units of days.}
\end{deluxetable}
\clearpage

\begin{deluxetable}{ccccccc}
\tabletypesize{\small}
\tablewidth{0pt}
\tablecaption{Fractional Accuracy of Velocity from Ground-Based Observations\label{tab_four}}
\tablehead{
\multicolumn{1}{c}{} & \multicolumn{3}{c}{$P =$ 3 days} & \multicolumn{3}{c}{$P =$ 1 year} \\
\hline
%\multicolumn{1}{c}{$R_p$} & \multicolumn{1}{c}{} & \multicolumn{1}{c}{} & \multicolumn{1}{c}{} & \multicolumn{1}{c}{} & \multicolumn{1}{c}{} & \multicolumn{1}{c}{} \\
\multicolumn{1}{c}{$R_p$ ($R_{\oplus}$)} & \multicolumn{1}{c}{$r'$} & \multicolumn{1}{c}{$i'$} & \multicolumn{1}{c}{$z'$} & \multicolumn{1}{c}{$r'$} & \multicolumn{1}{c}{$i'$} & \multicolumn{1}{c}{$z'$}}
\startdata
4     & 0.573    & 0.570    & 0.552   & 0.329  & 0.302  & 0.277 \\
11    & 0.0809   & 0.0780   & 0.0746  & 0.0459 & 0.0436 & 0.0416 \\
\enddata
\tablenotetext{a}{The accuracies were determined based on simulated ground-based photometry for a non-central transit ($b = 0.5$) of a planet orbiting a solar-like star.  Gaussian photometric noise of 1.7 mmag was added to each 1 minute observation based on the typical precision obtained in the Transit Light Curve project observations of TrES-1 \citep{winn2007}.  Linear ($u_1$) and quadratic ($u_2$) limb darkening coefficients were determined from ATLAS models \citep{claret2004}.  The coefficients ($u_1, u_2$) used for generating simulated light curves in the $r'$-, $i'$-, and $z'$- bands are (0.433, 0.30215), (0.3183, 0.3375), and (0.22785, 0.3573).}
\end{deluxetable}
\clearpage


\end{document}